\documentclass[12pt]{article}
\usepackage{amsmath,amssymb,latexsym,amsthm}

\def\R{\mathbb R}
\DeclareMathOperator{\St}{St}

\theoremstyle{remark}
\newtheorem*{remark}{Remark}
\theoremstyle{plain}
\newtheorem{proposition}{Proposition}
\newtheorem*{theorem}{Theorem}

\begin{document}
\title{
\vspace{1cm} {\bf Hierarchy of space-time structures, Boltzmann equation, and functional mechanics}
 }
\author{A.\,S.~Trushechkin\bigskip
 \\
{\it  Steklov Mathematical Institute, Russian Academy of Sciences}
\\ {\it Gubkina St. 8, 119991 Moscow, Russia}\medskip\\
{\it National Research Nuclear University ``MEPhI''}\\
{\it Kashirskoe highway 31, 115409 Moscow, Russia}\bigskip
\\ email:\:\texttt{trushechkin@mi.ras.ru}}

\date {}
\maketitle

\begin{abstract}
In this report we discuss the organization of different levels of nature and the corresponding space-time structures by the consideration of a particular problem of time irreversibility. The fundamental time irreversibility problem consists in the following: how to reconcile the time-reversible microscopic dynamics and the irreversible macroscopic one.

The recently proposed functional formulation of mechanics is aimed to solve this problem. The basic concept of this formulation is not a material point and a trajectory, like in the traditional formulation of mechanics, but a probability  density function. Even if we deal with a single particle (not with an ensemble of particles), we describe its state as a probability density function. We justify this approach using measurement theory.

A particular problem in the framework of the irreversibility problem is the derivation of the Boltzmann kinetic equation from the equations of microscopic dynamics. We propose a procedure for obtaining the Boltzmann equation from the Liouville equation based on the BBGKY hierarchy, the recently proposed functional formulation of classical mechanics, and the distinguishing between two scales of space-time, i.e., macro- and microscale. The notion of a space-time structure is introduced. It takes into account not only the space-time itself (i.e., a pseudo-Riemannian manifold), but also a characteristic length and time. The space-time structures form a hierarchy in sense that the initial values for the processes on the microscopic space-time structure (interactions of the particles) are assigned from the processes on the macroscopic one (kinetic phenomena).

\end{abstract}

\section{Introduction. Irreversibility problem}

The laws of physics are formulated in terms of differential equations. Every physical law is a mathematical model of the corresponding level of nature. For the last centuries theoretical and mathematical physics made the great success in the revelation of these laws as well as in the consequent technical progress. However, in order to have a holistic picture of the world, we should not only describe its separate levels, but also understand, how these descriptions relate to each other.

One conception is reductionism. It claims that the properties of a system can, in principle, be derived from the properties of its parts. For example, the dynamics of a gas can be derived from the dynamics of its atoms or molecules. But it seems that the reductionism does not work: a system as a whole has some properties which cannot be understood from the studying of its parts. For example, the temperature characterise the gas a whole and cannot be understood from the level of separate atoms or molecules.

Another property of this type is the property of irreversibility of macroscopic systems in time. Laws of the microscopic motion, given by the Newton, Hamilton, Schr\"odinger equations, quantum field theory equations etc., are time-symmetric, i.e., all physical processes can go to both directions of time. On the contrary, the laws of macroscopic motion, given by the Boltzmann, Navier-Stocks, other kinetic and hydrodynamic equations, the second law of thermodynamics etc., have a distinguished direction of time.

Let us consider this problem more mathematically. The Newtonian formulation of classical mechanics based on the notions of point and trajectory in the phase space. The fundamental law of dynamics in this formulation is given by the Hamilton system of equations. Consider the system of $N$ identical particles. Then the Hamilton equations have the following form:

\begin{equation}\label{EqHamiltonEqs}
\dot q_i=\frac{\partial H}{\partial p_i},\quad\dot p_i=\frac{\partial H}{\partial q_i},
\quad i=1,2,\ldots,N.\end{equation}
Here $q_i=q_i(t)\in G\subset\mathbb R^3$ is the position of the $i$-th particle at time $t$, $p_i=p_i(t)\in\mathbb R^3$ is the momentum of the $i$-th particle at time $t$, $H=H(x_1,\ldots,x_N)$ is a Hamiltonian, $x_i=(q_i,p_i)$. The following proposition holds:

\begin{proposition}
Let $(q(t),p(t))$ be a solution of the Hamiltonian system of equations (\ref{EqHamiltonEqs}), $q(t)=(q_1(t),\ldots,q_N(t))$, $p(t)=(p_1(t),\ldots,p_N(t))$. Let $H(q,p)=H(q,-p)$. Then $(\widetilde q(t),\widetilde p(t))=(q(-t),-p(-t))$ is also a solution of the Hamiltonian system.
\end{proposition}

The proof is straightforward. This proposition is a mathematical expression of the time-reversibility of the classical dynamics of particles, since the most of the fundamental Hamiltonians satisfy the condition $H(q,p)=H(q,-p)$.

We suppose that $G$ is a bounded region in $\mathbb R^3$ with the volume $V$ and a smooth boundary $\partial G$. We will consider the Hamiltonians of the form

\begin{equation}\label{EqH}
H=\sum_{i=1}^N\frac{p_i^2}{2m}+
\sum_{\begin{smallmatrix}i,j=1\\i>j\end{smallmatrix}}^N\Phi(\frac{|q_i-q_j|}\mu)+\sum_{i=1}^NU(q_i).\end{equation}
Here $\Phi(r)$ is an interaction potential of the particles, $U(q)$ is an external potential, $m>0$ is the mass of a particle. We assume that $\Phi(r)$ is continuously differentiable, monotonically decrease (which corresponds to the repelling force) on $(0,+\infty)$, $\Phi(r)\to0$ as $r\to+\infty$, $\Phi(r)\to+\infty$ as $r\to0$. $U(q)$ is continuously differentiable in $G$, $U(q)\to+\infty$ as $q\to\partial G$ (i.e., the external potential does not allow the particles to leave the domain $G$). $\mu>0$ is a dimensionless scale parameter which we will use later.

Consider now a gas from the point of view of kinetic theory. In this case the gas is described in terms of a single-particle density function $f(q,p,t)$. Let it be normalized on the volume $V$ (i.e., $f(q,p,t)/V$ is a probability density function), $q\in G$, $p\in\R^3$. One of the fundamental equations of the kinetic theory is the Boltzmann equation:

\begin{equation}\label{EqBoltz}
\frac{\partial f}{\partial t}=-\frac{p}m\frac{\partial f}{\partial q}+\frac{\partial U}{\partial q}\frac{\partial f}{\partial p}+\St f,\end{equation}
\begin{equation*}
\St f=n\int_{\R^2\times\R^3}\frac{|p-p_1|}m
[f(q,p',t)f(q,p'_1,t)-f(q,p,t)f(q,p_1,t)]d\sigma dp_1,\end{equation*}
where $n>0$ is the concentration of the particles (the number of particles in the unit of volume), $d\sigma=rdrd\varphi$. Also $p'$ and $p'_1$ are the momenta that will have two particles long after the collision, if they had the momenta $p$ and $p_1$ long before the collision with the impact parameter of the collision (scattering) $r$ and the polar angle $\varphi$ (so, $(r,\varphi)\in\R^2$ are polar coordinates on the plane perpendicular to the relative velocity vector $(p-p_1)/m$). Thus, $p'=p'(p,p_1,r,\varphi)$, $p'_1=p'_1(p,p_1,r,\varphi)$, the dependence is defined by the two-particle Hamiltonian without the external potential $H_2=\frac{p_1^2}{2m}+\frac{p_2^2}{2m}+\Phi(\frac{|q_1-q_2|}\mu)$. The expression $\St f$ is called the collision integral.

This is an important nonlinear equation which describes the relaxation of the function $f$ to the Maxwell distribution. One of the properties which can be easily proved is the so-called Boltzmann $H$-theorem:

\begin{proposition}[$H$-theorem]
Let f(q,p,t) be a solution of the Boltzmann equation (\ref{EqBoltz}) and the quantity
$$\mathcal H(t)=\int_{\Omega_V}f(q,p,t)\ln f(q,p,t)\,dqdp,$$
$\Omega_V=G\times\R^3$, be well-defined (i.e., the integral converges). Then $\frac{d\mathcal H}{dt}\geq0$.
\end{proposition}

This proposition states the entropy growth (the quantity $S(t)=-\mathcal H(t)$ can be regarded as the entropy of the gas) and, hence, the irreversible character of the gas dynamics.

Thus, we have obtained two contradictory conclusions: if we consider the gas as a whole (in terms of the single-particle distribution and the Boltzmann equation), the dynamics is irreversible in time, while if we consider the gas as a system of a finite number $N$ of particles (in terms of the point in the $N$-particles' phase space $\Omega_V^N$ and the Hamiltonian equations), the dynamics is reversible.

It seems that the reductionism does not work, the time irreversibility is a property of macrosystems which cannot be reduced to the microscopic level. So, the problem about another type of relation between the different levels of description arises.

\section{Functional formulation of classical mechanics and measurement theory}

Recently, a new formulation of classical mechanics called functional formulation was proposed by I.\,V.~Volovich \cite{VolFuncMech,VolRand} and is aimed to solve the irreversibility problem. This formulation is based not on the notions of a material point and a trajectory, but on the notion of a probability density function for a finite number of particles. Consider again a system of $N$ identical particles. In the functional formulation the state of this system at time $t$ is given by a probability density function $\rho(x_1,\ldots,x_N,t)$. The fundamental law of motion in this formulation is given by the Liouville equation
$$\frac{\partial\rho}{\partial t}=\{H,\rho\},$$
where $\{\cdot,\cdot\}$ are the Poisson brackets.

Formally, the Liouville equation is time-symmetric. But the solutions of the Liouville equation have the property of delocalization. For example, if we have one particle in a box with hard walls (i.e., $N=1$, $U(q)=0$ if $q\in G$, $U(q)=+\infty$ if $q\notin G$), then the spatial probability distribution $\sigma(q,t)=\int_{\R}\rho(q,p,t)dp$ tends to the uniform distribution in $G$. So, initially spatially well localized particle diffuses and fill the whole of the aivailablle volume. This delocalization property can be regarded as the irreversibility. So, the point of the functional formulation of mechanics is introducing the irreversibility on the microscopic level by the assignment of the holistic statistical properties to a single particle.

The argument of considering the probability density function rather than the material point as a fundamental object of mechanics is given by the measurement theory. Since the real (more precisely, irrational) numbers are unobservable, the notions of a material point and a trajectory, based on the concept of a real number, have not a direct physical meaning. The result of every measurement is a rational number (it is worthwhile to note that the justification of the $p$-adic and adelic mathematical physics also starts from this fact \cite{DragovichReview}). Moreover, every measurement has an error. So, we never know a state of the system as a point in the phase space. Every experimenter knows a state as a rational number and an error represented as a confidence interval and a confidence probability. This situation is naturally described by a probability density function.

For example, let an experimenter performs some measurements of some observable $\mathbb X$ and represent the results of the measurements in the form
$$x\in(x_0-\Delta_{0,997},x_0+\Delta_{0,997}),\quad \text{the confidence probability is } 0,997.$$
Here $x$ is an ``actual'' value of $\mathbb X$, i.e., he do not know the value of $\mathbb X$, but we know that it lies in the confidence interval $(x_0-\Delta_{0,997},x_0+\Delta_{0,997})$ with the probability 0,997. Alternatively, he can represent the result of the same measurements in the same form with some another confidence probability $\gamma$ and the corresponding $\Delta_\gamma$ (of course, $\Delta_\gamma$ increases as $\gamma$ increases). In fact, the experimenter deals with a probability density function. For example, under certain conditions this probability density function may have the form
$$\rho(x)=\frac1{\sqrt{2\pi}\sigma}e^{-\frac{(x-x_0)^2}{2\sigma^2}},$$
where $\sigma=\Delta_{0,997}/3=\Delta_{0,683}$ \cite{Gnedenko}.

Thus, the experimenter knows the state of the system not as a point in a phase space, but as a family of confidence intervals with the corresponding confidence probabilities, or, equivalently, as a probability distribution in the phase space. One can say that since every measurement has an error and the theory of measurement error is based on mathematical statistics and probability theory, a description of the state of any classical physical system in terms of probability distributions is more natural and direct from the experimental point of view.

Assigning a certain values for the physical quantities, as it is accepted in the traditional view on mechanics, is an abstraction and idealization. As every idealization, it works in some situations, but in other situations it does not. Of course, we can not neglect the measurement errors, when we deal with large times, because the errors grow with the time due to the discussed delocalization effect. Such large time asymptotics as thermalization or recurrence (almost periodicity) are often discussed with the relation to the irreversibility problem. So, if we examine the time irreversibility problem, we must consider mechanics in its functional, not the traditional, formulation.

It should be noted that the use of a real-valued probability density function does not contradict to the rational-valued measurement results. The probability density function is not a directly observable value, rather it is constructed from the directly observable results of measurements. See \cite{TrVolFuncRat} for the detailed description of the construction of the probability density function starting from the measurement results. See also \cite{TrushTMF10} for the discussing the functional dynamics of a system under often measurements. Since the probability density function is not directly observable, it can be real-valued. Like the concept of a material point, the concept of a probability density function is also a kind of idealization. But this idealization works in a more general case.

\section{Derivation of the Boltzmann equation and the hierarchy of space-time structures}

One of the problems of the framework of the irreversibility problem is the derivation of the Boltzmann equation from the Liouville equation. Two different derivations have been got by Bogolyubov \cite{Bogol} and Lanford \cite{Lanford}. The main drawbacks of the Bogolyubov's derivation are the number of the additional assumptions and the divergencies in the high order corrections to the Boltzmann equation \cite{Bogol77}. The main drawback of the Lanford's derivation is a small time on which the validity of the Boltzmann equation is proved. This time appears to be approximately $1/5$ of the mean free time \cite{Spohn}. On the contrary, the Boltzmann equation is aimed to explain the dynamics of the density function and its convergence to the Maxwell distribution on the times much greater than the mean free path. In this section we present an alternative derivation of the Boltzmann equation from the Liouville equation.

Both Bogolyubov and Lanford start with the Cauchy problem for the Liouville equation
\begin{equation}\label{EqLiouvCauchy}
\left\lbrace\begin{aligned}
&\frac{\partial\rho}{\partial t}=\{H,\rho\},\\
&\rho(x_1,\ldots,x_N,0)=\rho^0(x_1,\ldots,x_N).
\end{aligned}\right.
\end{equation}
As we have discussed in the previous section, we construct the initial probability density function $\rho^0$ from the results of measurements of each particle's position and momentum. But practically, when we consider the kinetic events, we do not measure the position and momentum of each of $N$ (which can be of order Avogadro number $10^{23}$) particles. In fact, there are two space-time scales. The first one is the scale of the particles interaction radius $r_0$ and the particle interaction time $r_0/{\bar u}$ (a microscale). Here $\bar u$ is the mean velocity of the particles. The second one is the scale of the mean free path $l$ and the mean free time $l/{\bar u}$ (a macroscale or a scale of kinetic phenomena). The Boltzmann equation is valid if $r_0\ll l$. We put $\mu=\frac{r_0}l\to0$. This parameter in the Hamiltonian (see (\ref{EqH})) means that particles interact on the scale much smaller than the considering scale of $l$.

In practice, when we consider the kinetic events, we have a measuring instrument which can capture the changes of the functions only on the macroscale. We can construct the one-particle probability density function $\rho^0_1(x_1)=\int_{\Omega_V^{N-1}}\rho^0 dx_2\ldots dx_N$ using this instrument (denote also $f^0_1=V\rho^0_1$). But we cannot construct the function $\rho^0$, because this function incorporates the information about the correlations of the particles on the distances of order $r_0$ which cannot be captured by our instrument.

So, the Cauchy problem (\ref{EqLiouvCauchy}) has not a direct physical meaning. Instead of the Cauchy problem we start with the following problem:

\begin{equation}\label{EqLiouvHierarchy}
\left\lbrace\begin{aligned}
&\frac{\partial f}{\partial t}=\{H,f\},\\
&S^{(2)}_{-\Delta t}[f_2(x_1,x_2,t-\Delta t)-f_1(x_1,t-\Delta t)f_1(x_2,t-\Delta t)]\to0\\
&\quad\text{(as $\mu\to0$, $N\to\infty$, $N\mu^2=const$, and $\Delta t\to0$, $\frac{\Delta t}\mu=\Delta\tau\to\infty$)},\\
&f_1(x_1,0)=f_1^0(x_1),\\
&\int_{\Omega_V^N} Hf\,dx_1\ldots dx_N<\infty,\\
&f(x_1,\ldots,x_N,t)=f(x_{i(1)},\ldots,x_{i(N)},t),\quad (i(1),\dots,i(N))=P(1,\ldots,N),
\end{aligned}\right.
\end{equation}
Here $f(x_1,\ldots,x_N,t)=V^N\rho(x_1,\ldots,x_N,t)$ (of course, $f$ depends also on $\mu$, since $\mu$ is a parameter in the Hamiltonian, the initial function $f^0_1$ also may depend on $\mu$); $$f_s(x_1,t)=V^{N-s}\int_{\Omega_V^{N-s}}f(x_1,\ldots,x_N,t)\,dx_{s+1}\ldots dx_N,$$ $s=1,2$; $S^{(2)}_t$ is the two-particle Hamiltonian flow, i.e., $S^{(2)}_t\varphi(x_1,x_2)=\varphi(x_{1t},x_{2t})$, where $(x_{1t},x_{2t})$ is the phase point $(x_1,x_2)$ moved along the flow on $t$; $P(1,\ldots,N)$ is a permutation of the numbers $1,\ldots,N$.

\begin{theorem}
Let the function $f(x_1,\ldots,x_N,t)$ satisfy the problem (\ref{EqLiouvHierarchy}). Let $\Phi(r)$ be monotonically decreasing function
and $\lim\limits_{r\to\infty}r^\gamma\Phi(r)=C\neq0,$ $\gamma>2$. Then the function $f_1(x_1,t)$
satisfies the Boltzmann equation (\ref{EqBoltz}) with the initial function $f^0_1$ in the limit $\mu\to0, N\to\infty, N\mu^2=const=nV$ (i.e., $\frac{\partial f_1(x,t)}{\partial t}$ converges to the right-hand side of (\ref{EqBoltz}) with $f_1$ instead of $f$ in every  point $x\in\Omega_V$ and $t\in\R$).
\end{theorem}

This is the main result of this report. The limit $\mu\to0$, $N\to\infty$, $N\mu^2\to const$ is called the Grad or Boltzmann--Grad limit \cite{Grad} and was used by Lanford. Bogolyubov used the thermodinamic limit $N\to\infty$, $V\to\infty$, $\frac NV=n=const$. There is also a derivation of Bogolyubov--Boltzmann-type equation for a finite number of particles (for example, for two particles), i.e., without the limit $N\to\infty$, based on the functional approach to mechanics \cite{VolBogol}.

Let us discuss the formulation of problem (\ref{EqLiouvHierarchy}). The third condition is the initial value for the single-particle function $f_1$. The fourth condition means the finiteness of energy, the fifth condition means the symmetry of the density function with respect to permutations (i.e., the particles are indistinguishable).

The most interesting and crucial is the second condition. The knowledge of the initial single-particle function $f^0_1(x_1)$ is not sufficient to get a unique solution for the single-particle function $f_1(x_1,t)$. Since, as we said above, the condition of the form $f(x_1,\ldots,x_N,0)=f^0(x_1,\ldots,x_N)$ has not a direct physical meaning, we must have some additional condition.

As we said above, there are two scales of space-time in this consideration: the microscale, related to the interaction of particles, and the macroscale, related to the kinetic phenomena. The two-particle function $f_2$ relates to the microscopic scale, because it incorporates the information about the pairwise correlations of the particles on the distances of  order $r_0$. The single-particle function $f_1$ relates to the macroscopic scale, since it does not incorporate the information about particles' interaction and the kinetic theory is expressed in terms of this function. The second condition in (\ref{EqLiouvHierarchy}) means that the initial value for the microscopic function $f_2$ are assigned by the macroscopic function $f_1$. So, instead of the specification of the initial microscopic function $f^0$, we specify only the initial macroscopic function $f^0_1$ and impose a condition on the microscopic function: in a certain sense it is subordinated to the macroscopic one (in sense that its initial values are assigned from the macroscale).

Traditionally space-time is defined as a manifold $M$ with a pseudo-Riemannian metrics $g$. In the simplest case this is $\R^4$ with the usual Lorentz metrics $\eta$ (the Minkowski space). However, when we speak about a space-time, we keep in mind some characteristic length and time. For example, in continuum mechanics the density is defined as $\frac{dm(r)}{dV}=\lim\limits_{\Delta V\to0}\frac{\Delta m(r)}{\Delta V}$, where $\Delta m(r)$ is the mass of the volume $\Delta V$ near the point $r\in\R^3$. In spite of the mathematical limit $\Delta V\to0$, i.e., the limit of arbitrarily small volumes, we keep in mind that $\Delta V$ is a macroscopic volume, i.e., it contains a large number of atoms or molecules (otherwise the definition is not valid).

So, always when we deal with some space-time, we keep in mind not only the space-time $(\R^4,\eta)$ itself, but also a characteristic length $L$, which tell us about the averaging scale (the characteristic time is given by $L/{\overline u}$, where $\overline u$ is the mean velocity). In our example we average over the molecular structure of the medium. Exactly this allows us to perform the limit $\Delta V\to0$. The triple $(\R^4,\eta,L)$ we will call the space-time structure.

In our problem of the derivation of the Boltzmann equation we also actually deal with two distinct space-time structures. The first one is the space-time structure $(G\times\R,\eta,l)$ of macroscopic (kinetic) phenomena with the coordinates $(q,t)$ and the characteristic length $l$ (the mean free path). The second one is the space-time $(\frac G\mu\times\R,\eta,r_0)$ of microscopic phenomena, like particle interactions, with the coordinates $(\xi,\tau)$ and the characteristic length $l$ (the particle interaction radius). These space-time structures are related to each other by the scale transformation
\begin{equation}\label{EqRescal}
\xi=\frac q\mu,\quad\tau=\frac t\mu,\quad\mu=\frac{r_0}l\to0.
\end{equation}
With such a transformation infinitesimal region of the macroscopic space-time may correspond to the infinitely large region of the microscopic space-time. Exactly this situation we can see in the second condition in (\ref{EqLiouvHierarchy}): $\Delta t\to0$, but $\Delta\tau=\frac{\Delta t}\mu\to\infty$. Of course, these two scales meet each other in the collision integral of the Boltzmann equation: a collision is considered as a point and momentary act on the macroscale, but it takes place on the infinite space during the infinite time on the microscale.

The great disparity between microscopic and macroscopic scales as one of the origins of the irreversible macroscopic behaviour was pointed out in \cite{KozTres,Lebowitz}. The used rescaling of the space-time (\ref{EqRescal}) is typical for the derivation of the Boltzmann equation from the Liouville equation \cite{Spohn,DobrSinaiSukhov}. In the case of lattice dynamics this rescaling was used in \cite{DudSpohn,Dud} for the derivation of kinetic and hydrodynamic-type equations. Our proposition is to introduce the subordination of different space-time structures expressed in the form of the second condition in (\ref{EqLiouvHierarchy}), which gives a new way of derivation of the Boltzmann equation.

One can say about a kind of hierarchy of the space-time structures: the initial values for the processes on the microscopic space-time structure (interactions of the particles) are assigned from the processes on the macroscopic space-time structure (kinetic phenomena). Note that the idea of the hierarchy of times (namely, the microscopic, kinetic and hydrodynamic times) in a slightly different sense was first proposed by Bogolyubov \cite{Bogol}.

\begin{remark}We can note that in our limiting process the overall mass of the gas $M=mN$ tends to infinity, since $m$ is constant and $N$ tends to infinity. However, we can also rescale the mass of a single particle in the way $m_\mu=\mu^2m$ and rescale the interaction potential in the way $\Phi_\mu(r)=\mu^2\Phi(r)$ as it was proposed by Grad \cite{Grad} (we substitute $m$ by $m_\mu$ and $\Phi(\frac{|q_i-q_j|}\mu)$ by $\Phi_\mu(\frac{|q_i-q_j|}\mu)$ in (\ref{EqH})). The Hamiltonian system with these additional rescalings can be easily reduced to the considered one. So, the gas is considered as an infinite number of negligible mass particles in the limit. This exactly corresponds to the macroscopic intuition.
\end{remark}

Finally, we would like to note that the functional approach can be useful in the mothod of molecular dynamics simulations. At present the molecular dynamics follows the Newtonian approach and simulates the movement of material points. On this way it is hard to obtain the properties of complex and, moreover, biological systems, which constitute the aim of the molecular dynamics. Furthermore, the problem of uncontrolled cumulative errors in numerical integration is known. The simulation of the motion in terms of the probability density function seems to be more appropriate for obtaining the properties of the complex and biological systems and more justified from the computational point of view. Also the present report suggests that the initial conditions for the probability density function should be chosen not in an arbitrary way, but rather in the form suggested by the system as a whole. In our case such conditions are given by (\ref{EqLiouvHierarchy}). Let us note that these conditions are statistical in essence, they are formulated for the probability distribution and cannot be reduced to the conditions on the initial positions and momenta of the individual particles.

\section{Conclusions}
In this report we have reviewed the irreversibility problem and the functional formulation of mechanics. This formulation is based on the concept of a probability density function. The fundamental equation of this formulation is the Liouville equation for the probability density function. The functional approach to mechanics is justified by the measurement theory. We have proposed a derivation of the Boltzmann equation from the Liouville equation based on some ideas of the functional formulation and the measurement theory. The main features of this derivation is distinguishing of two scales of space-time (micro- and macroscopic or kinetic) and subordination of the processes on the microscopic space-time structure to the processes on the macroscopic one.

According to the traditional paradigm of the mathematical physics, the dynamics is completely determined if we know the law of motion, i.e., a differential equation, and the initial values for it. However, the initial values themselfs are understood as something external to the equations of mathematical physics (``As regards the present state of the world\ldots the laws of nature are entirely silent'' \cite{Wigner}). We propose another picture: the initial values for a given level of nature are assigned from the higher level.

So, instead of the reductionism, which claims the reducibility of all levels of the nature to the most microscopic level, we propose a hierarchical picture of the world: the lower levels of the nature are subordinated to the higher ones.

\section{Acknowledgments}

The author is grateful to Prof. I.\,V.~Volovich for helpful discussions and remarks. This work was partially supported by the Russian Foundation of Basic Research (projects 08-01-00727-a and 09-01-12161-ofi-m), the grant of the President of the Russian Federation (project NSh-7675.2010.1), and the Division of Mathematics of the Russian Academy of Sciences.

\end{document}